\documentclass[%
 reprint,
 amsmath,amssymb,
 showkeys
floatfix,
]{revtex4-2}

\usepackage[T1]{fontenc}
\usepackage{graphicx}
\usepackage{dcolumn}
\usepackage{bm}

\usepackage[colorlinks=true,linkcolor=blue,urlcolor=blue,citecolor=blue,pdfusetitle]{hyperref}
\usepackage[english]{babel}
\usepackage{float}
\usepackage{afterpage}
\usepackage{amsmath}
\usepackage[caption = false]{subfig}
\usepackage{graphicx,epstopdf}
\usepackage{blindtext}
\usepackage[table,xcdraw,dvipsnames]{xcolor}
\usepackage{lipsum}
\usepackage{amsfonts}
\usepackage{bbm}
\usepackage{amssymb}
\usepackage{enumerate}
\usepackage{xcolor}
\usepackage{latexsym}
\usepackage{physics}
\usepackage{times,txfonts}
\usepackage{soul}
\usepackage[normalem]{ulem}
\usepackage{indentfirst}

\begin{document}

\title{Interaction-Driven Photocurrent Enhancement in a Quantum Photocell with Opposite Transient and Steady-State Responses}
\author{Abbas Shokri} 
\email{abbasshokri4341@gmail.com}
\affiliation{Physics Department, Azarbaijan Shahid Madani University, Tabriz, Iran}

\author{Esfandyar Faizi} 
\email{efaizi.quantum@gmail.com}
\affiliation{Physics Department, Azarbaijan Shahid Madani University, Tabriz, Iran}

\author{Mohammad B. Arjmandi}
\email{arjmandi94@gmail.com}
\affiliation{ Wilczek Quantum Center, Shanghai Institute for Advanced Studies, University of Science and Technology of China, Shanghai 201315, China.}

\date{\today}

\begin{abstract}
\section*{Abstract}
We study photocurrent control in a quantum photocell with interacting qubit–qutrit donors coupled to an acceptor. We show that stronger Heisenberg exchange coupling and anisotropy increase the transient photocurrent but reduce its steady-state value, whereas the Dzyaloshinskii–Moriya interaction favors steady-state output at the expense of transient enhancement. This reversal highlights the need to distinguish rapid current generation from sustained operation when tuning quantum photovoltaic devices. Analysis of donor purity further reveals that the qutrit’s local purity tracks the photocurrent trends in both regimes. These results demonstrate that the effects of donor interactions depend crucially on the timescale of operation, suggesting a strategy for tailoring the response of hybrid quantum photocells to different operating regimes.
\end{abstract}
\maketitle
\section{\label{sec:level1}Introduction}
Photovoltaic cells convert solar radiation into electrical energy through photon absorption and the subsequent generation of electron-hole pairs across a semiconductor bandgap~\cite{r1,r2,r3,r4,r5}. As one of the most promising pillars of sustainable energy, these devices have undergone remarkable advances in efficiency and cost reduction~\cite{r6,r7,r8,r9,r10}. Nevertheless, their performance remains fundamentally constrained by thermodynamic and quantum-mechanical limits that impose intrinsic ceilings on energy conversion~\cite{r11,r12,r13}.

The transition from classical to quantum photovoltaics marks a conceptual departure from band-structure engineering toward discrete-level architectures. Quantum photocells exploit systems with quantized energy levels, enabling precise control over absorption, transport, and energy storage~\cite{r14,r15,r16,r17,r18,r19,r20,r21,r22,r23,r24,r25,r26}. Such systems are commonly realized using quantum dots~\cite{r14,r15,r16,r17}, quantum wells~\cite{r18,r19}, or molecular chromophores~\cite{r20,r21}. In these platforms, quantum coherence can be harnessed to overcome classical limitations and enhance photovoltaic performance~\cite{r22,r23,r24}.

The quantum features like coherence and entanglement have already opened new frontiers across diverse technological domains, including quantum metrology~\cite{r25,r26}, quantum sensing~\cite{r27,r28,r29}, quantum teleportation~\cite{r30,r31}, photonic quantum information processing~\cite{r32,r33}, quantum key distribution~\cite{r34,r35,r36}, and quantum computation~\cite{r37,r38}. Moreover, quantum batteries~\cite{r39,r40,r41,r42,r43,r44,r45}—which rely on discrete energy-level structures for energy storage—share some conceptual similarities with quantum photocells. In both systems, energy is stored in excited states, and key performance indicators such as storage capacity, excited-state lifetime, and resilience against decoherence play a central role.

Despite these prospects, the primary impediment to achieving ultrahigh efficiencies remains the Shockley–Queisser limit—a theoretical bound formulated over six decades ago on the basis of detailed balance~\cite{r8}. This limit sets the maximum conversion efficiency of a single-junction solar cell at approximately 33\%, dictated by three fundamental loss channels: transmission of sub-bandgap photons, thermalization of hot carriers, and radiative recombination of electron-hole pairs. Among these, electron-hole recombination represents one of the most fundamental loss mechanisms. In this process, an excited electron recombines with its hole before contributing to the external circuit, releasing energy either as a photon (radiative recombination) or through phonon emission (non-radiative recombination)~\cite{r13,r24}.

In this context, quantum effects have emerged as effective tools for mitigating recombination and enhancing efficiency. It has been shown that coherent superposition of energy levels, induced by an external drive, can significantly suppress radiative recombination~\cite{r46}. Furthermore, environment-induced coherence—specifically Fano coherence—has been demonstrated to provide similar benefits without the need for external driving~\cite{r47,r48}. Studies on interacting donor systems, ranging from two, three-donor to N-donor architectures, have shown that excitonic coupling and quantum interference between donor states can enhance photocurrent and improve overall conversion efficiency~\cite{r49,r50,r51}. In a complementary approach, the use of photon pulses as an energy source has been shown to replace the hot thermal reservoir and enable efficient operation with a single cold bath~\cite{r24}. Similarly, multi-level quantum dot models have demonstrated that extending the Hilbert space can enhance photon absorption and suppress radiative recombination, leading to improved quantum yields~\cite{r52}.

Although most studies on quantum photocells have focused on the steady-state regime, the transient regime is also of significant practical importance. This regime is crucial for applications such as ultrafast optoelectronic devices, quantum information processing, and photodetectors, where the response speed and the charge extracted within a short time window are decisive. On the other hand, the steady-state regime characterizes the long-term performance and stability of the device and is essential for applications requiring a stable and continuous current. Therefore, examining both the transient and steady-state regimes can provide a more complete understanding of the photocell performance.

In this work, a hybrid quantum photocell composed of a qubit-qutrit donor is introduced, coupled via XXZ and Dzyaloshinskii–Moriya (DM) interactions~\cite{r53}, which in turn can lead electronic excitation into an acceptor. The open-system dynamics is modeled via a Lindblad master equation, and the photocurrent, voltage, power, and local purity of the donors are computed in both transient and steady-state regimes. The results show that increasing \(J\) and \(\kappa\) enhances the transient current while suppressing the steady-state current, whereas \(D_z\) exhibits an opposite trend in these two regimes. Purity analysis further shows that the qutrit’s local purity follows the current in both regimes, while the qubit’s purity shows no consistent correlation with the current, and its trend reverses between the transient and steady-state regimes. These findings establish hybrid qubit–qutrit architectures as a suitable platform for designing tunable quantum photovoltaic devices.

The remainder of this paper is organized as follows. Section~\ref{sec:level2} introduces the basic structure of a two-level donor–acceptor quantum photocell and discusses its elementary operating principles. Section~\ref{sec:level3} defines the key figures of merit—photocurrent, voltage, output power, and purity. The complete model, including the Hamiltonian of the qubit–qutrit donors and the Lindblad master equation, is presented in Section~\ref{sec:level4}. Finally, Section~\ref{sec:level5} presents and discusses the numerical results, and Section~\ref{sec:level6} concludes the paper with a summary and outlook.

\section{\label{sec:level2}Quantum Photocell: Basic Structure and Operating Principles}

Before proceeding to our central model—comprising two structurally distinct donors, a qubit and a qutrit, coupled through XXZ and DM interactions—it is instructive to first consider the simplest quantum photocell, in order to illustrate its basic operating mechanism. This preliminary exposition serves to establish the fundamental physical processes and the operational cycle in a transparent and physically intuitive manner, free from the technical complexities that accompany the full model. In its most elemental form, a quantum photocell consists of two essential building blocks: a donor, which captures incident photons, and an acceptor, which subsequently separates and extracts the photoexcited charge carriers, thereby enabling the generation of an external photocurrent. Unlike conventional solar cells, which rely on continuous energy bands in semiconductors, quantum photocells exploit discrete energy levels of their donor systems. This discrete structure enables coherent control over absorption, charge transfer, and recombination, offering potential advantages over classical counterparts.
\begin{figure}[!ht] 
\centering
\includegraphics[width=0.7\columnwidth]{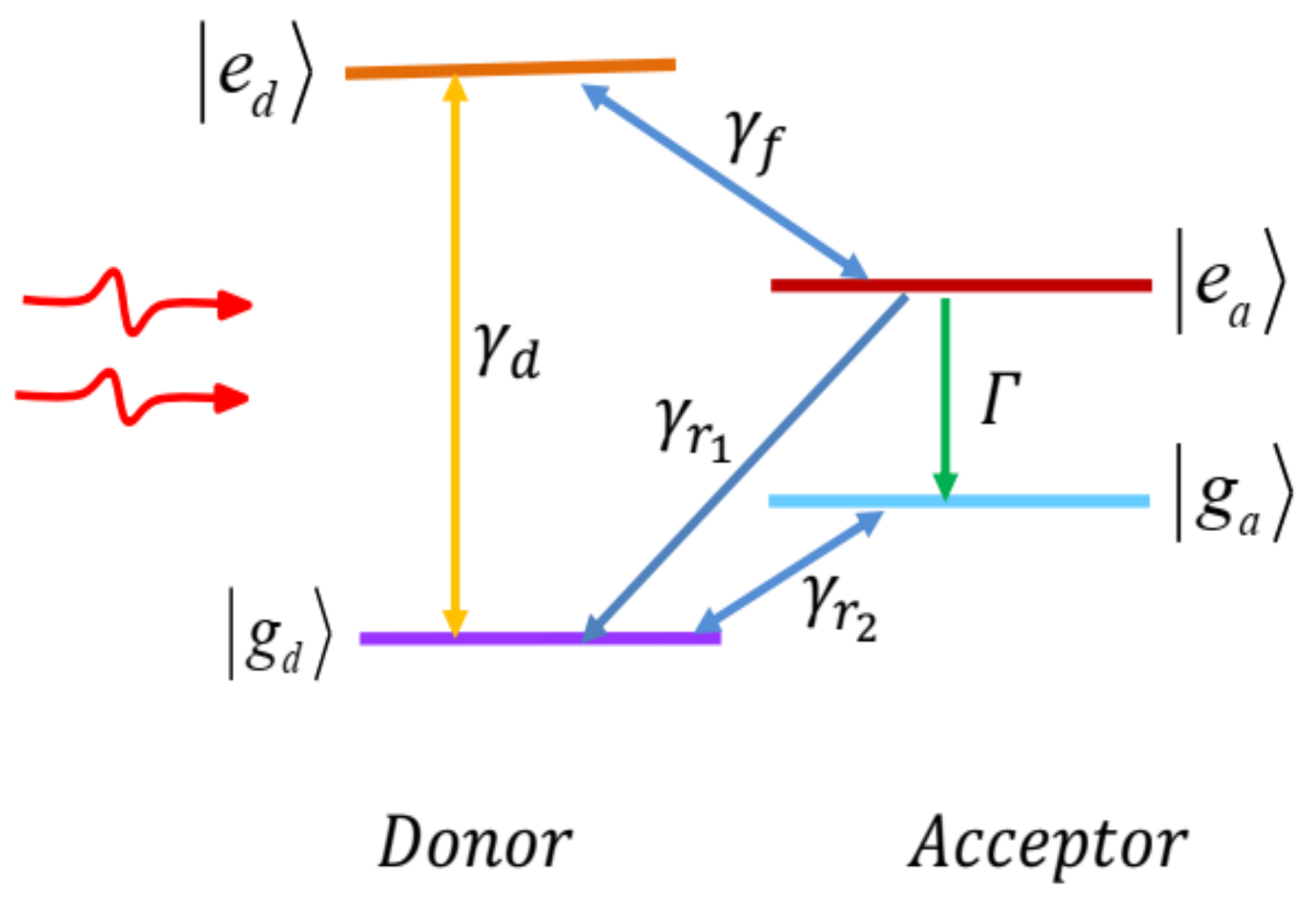}
\caption{Schematic energy-level diagram of the single-donor quantum photocell. The donor ($|g_d\rangle \leftrightarrow |e_d\rangle$) transfers the excited electron to the acceptor ($|e_a\rangle$) via a phonon-assisted process ($\gamma_f$), followed by decay through the external load ($\Gamma$) and photocurrent generation. Competing loss channels include direct recombination $\gamma_{r_1}$ and ground-state recombination $\gamma_{r_2}$.}
\label{f1}
\end{figure}

This single-donor model, illustrated in Figure~\ref{f1}, is based on the standard quantum photocell framework introduced in~\cite{r24}, in which both the donor and the acceptor are modeled as two-level systems (qubits). The operational cycle begins with photon absorption by the donor, where the electron transitions from the ground state \(\ket{g_d}\) to the excited state \(\ket{e_d}\), characterized by the rate \(\gamma_d\), which accounts for both emission and absorption. From this excited state, the electron follows two competing pathways. The desired path is the phonon-assisted move to the acceptor, denoted by the rate \(\gamma_f\), which brings the electron to the acceptor excited state \(\ket{e_a}\). From there, it decays to the external load at rate \(\Gamma\), producing useful current. Competing with this beneficial pathway is a radiative loss channel: direct recombination from the acceptor to the donor, \(\gamma_{r_1}\), which returns the electron from \(\ket{e_a}\) to \(\ket{g_d}\) without generating work. Finally, the cycle is closed via ground-state recombination \(\gamma_{r_2}\), corresponding to the transition \(\ket{g_a} \to \ket{g_d}\), which restores the system to its initial state.

The performance of a quantum photocell is thus governed by the competition between these beneficial and dissipative processes. The photon absorption rate depends on the incident flux and the donor's dipole coupling strength, while transfer and recombination are mediated by the phonon environment. An efficient cell is one in which the transfer rate to the acceptor dominates over the recombination rates, ensuring that most excited electrons reach the external load before being lost. In the following sections, this simple framework is built upon to introduce quantitative performance metrics—current, voltage, power, and purity—and then the model is extended to include a second donor with qutrit structure. This extended architecture, incorporating inter-donor coupling and quantum interference, allows new parametric dependencies to be explored and the output behavior to be analyzed across different operational regimes.

\section{\label{sec:level3}Figures of Merit for Quantum Photocell Performance}

To evaluate the performance of a quantum photocell, several measurable quantities are employed, each describing a specific aspect of the system's behavior. In this section, three standard figures of merit—photocurrent, voltage, and output power—that are commonly used to evaluate photocell performance are introduced. In addition, the analysis is extended by examining the local purity of the donors to assess how it is correlated with photocell performance.

\subsection{Photocurrent}

The electric current generated by the photocell arises from the decay of the acceptor excited state to its ground state via coupling to an external load at rate \(\Gamma\). The photocurrent is then defined as~\cite{r13}:
\begin{equation}
I = e \, \Gamma \, \langle e_a | \rho | e_a \rangle,
\label{eq:current}
\end{equation}
where \(e\) is the elementary charge, \(\Gamma\) is the decay rate to the load, and \(\langle e_a | \rho | e_a \rangle\) is the population of the acceptor excited state. Thus, the current is directly proportional to the occupation probability of the state \(\ket{e_a}\).

\subsection{Output Voltage}

The voltage generated by the system is determined by the energy gap of the acceptor levels and the population ratio of its states. It is given by~\cite{r13}:
\begin{equation}
V = \frac{1}{e} \left[ (E_{e_a} - E_{g_a}) + k_B T_{\text{pn}} \ln\left( \frac{\langle e_a | \rho | e_a \rangle}{\langle g_a | \rho | g_a \rangle} \right) \right],
\label{eq:voltage}
\end{equation}
where \(E_{e_a} - E_{g_a}\) is the energy spacing between the ground and excited levels of the acceptor, \(k_B\) is the Boltzmann constant, and \(T_{\text{pn}}\) is the temperature of the phonon reservoir. The logarithmic term accounts for the thermal correction arising from the population imbalance.

\subsection{Output Power}

The output power of the system is the product of current and voltage~\cite{r13}:
\begin{equation}
P = I \cdot V.
\label{eq:power}
\end{equation}
This quantity provides a direct measure of the photocell's capability as an electrical power generator and is typically evaluated at the maximum power point as the main performance metric.

\subsection{Purity}

Purity measures the lack of mixedness of a quantum state and is defined as~\cite{r46}:
\begin{equation}
\mathcal{P}(\rho) = \operatorname{Tr}\left( \rho^2 \right).
\label{eq4}
\end{equation}
The purity ranges between \(1/d\) for a maximally mixed state and \(1\) for a pure state, where \(d\) is the Hilbert space dimension. Higher purity indicates a state closer to pure, while lower purity indicates greater mixedness. In this work, purity is computed locally for each donor subsystem—both the qubit and the qutrit—by taking the partial trace over the remaining degrees of freedom. This allows us to assess the local mixedness of each donor and to correlate this quantity with the output photocurrent.

\section{\label{sec:level4}Model}

\begin{figure}[!ht] 
\centering
\includegraphics[width=0.9\columnwidth]{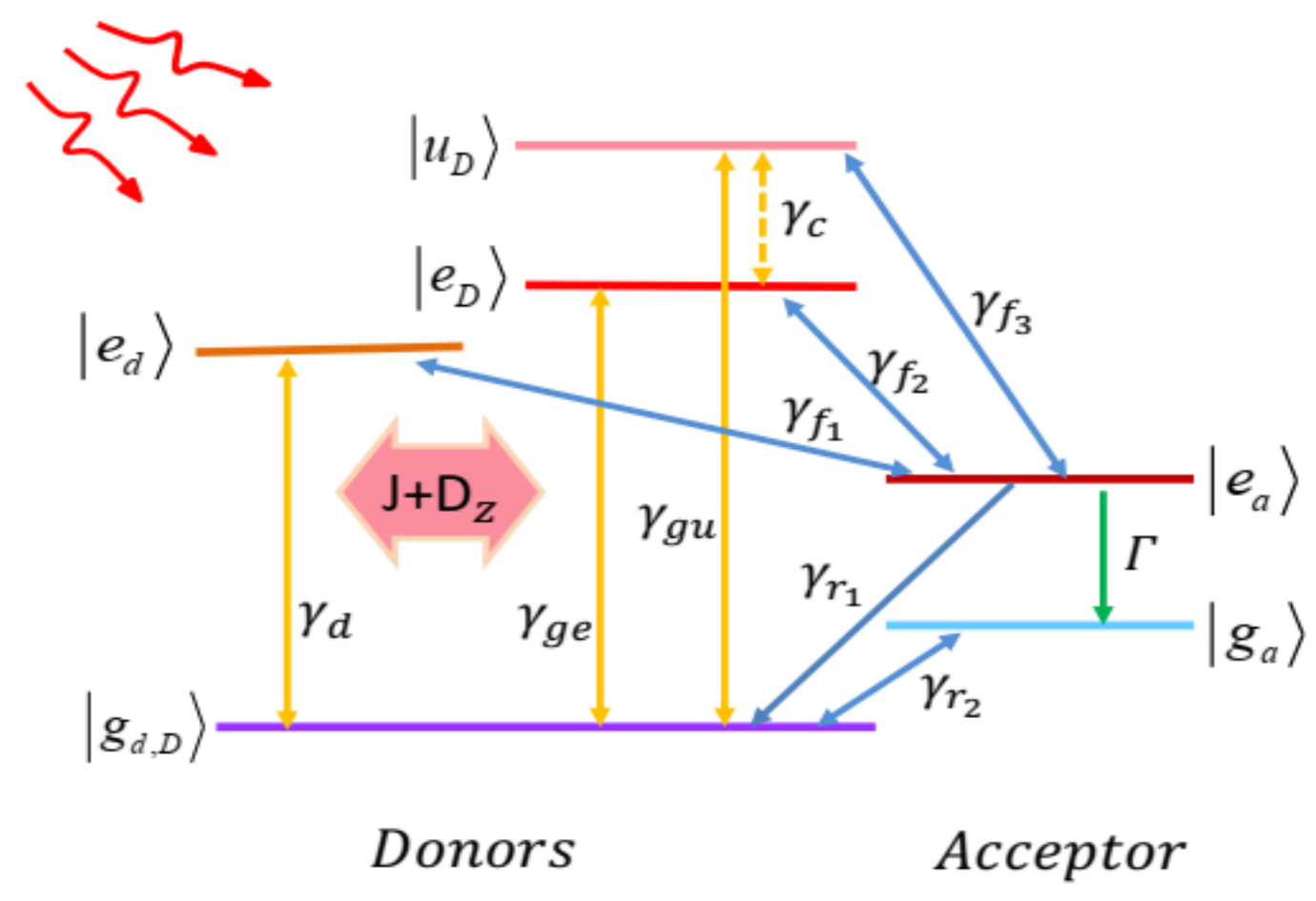}
\caption{Schematic energy-level diagram of the qubit–qutrit quantum photocell. The system comprises a qubit donor ($\ket{g_d},\ket{e_d}$), a qutrit donor ($\ket{g_D}, \ket{e_D},\ket{u_D}$), and a qubit acceptor ($\ket{g_a},\ket{e_a}$), with inter-donor coupling $J + D_z$. Orange arrows show photonic transitions ($\gamma_d$, $\gamma_{ge}$, $\gamma_{gu}$, and the cross-term rates $\gamma_x$ and $\gamma_y$, collectively denoted as $\gamma_c$), blue arrows denote phonon-assisted processes ($\gamma_{f_1},\gamma_{f_2},\gamma_{f_3}$ for forward transfer; $\gamma_{r_1},\gamma_{r_2}$ for recombination losses), and the green arrow indicates load extraction at rate $\Gamma$.}
\label{f2}
\end{figure}

Having established the basic operating principles of a single-donor quantum photocell, we now introduce the full model that constitutes the core framework of this work. This architecture, schematically depicted in Figure~\ref{f2}, comprises two donor subsystems of distinct nature: a qubit and a qutrit, along with an acceptor, which is typically modeled as a two-level system, forming a composite photovoltaic platform. In contrast to conventional qubit-only designs, the inclusion of a qutrit as the second donor extends the Hilbert space, allowing access to additional transition channels and coherence pathways, along with the donors coupling that may enhance the photovoltaic performance.

The first donor is a qubit with states \(\ket{g_d}\) (ground) and \(\ket{e_d}\) (excited), while the second donor is a qutrit with states \(\ket{g_D}\) (ground), \(\ket{e_D}\) (first excited), and \(\ket{u_D}\) (second excited). The acceptor is also a two-level system with states \(\ket{g_a}\) and \(\ket{e_a}\). The ground states of both donors are assumed to have the same energy, whereas the acceptor ground state is higher in energy, and its excited state lies below the donor excited level.

The total Hamiltonian of the system is composed of three contributions: the self-energy of the individual levels, the XXZ exchange interaction and the Dzyaloshinskii–Moriya (DM) interaction between the donors. The self-energy term represents the bare energies of each level and is expressed as:
\begin{equation}
H_{\text{self}} = H_d + H_D + H_a,
\end{equation}
where
\begin{align}
H_d &= E_{g_d} \ket{g_d}\bra{g_d} + E_{e_d} \ket{e_d}\bra{e_d},\\
H_D &= E_{g_D} \ket{g_D}\bra{g_D} + E_{e_D} \ket{e_D}\bra{e_D} + E_{u_D} \ket{u_D}\bra{u_D},\\
H_a &= E_{g_a} \ket{g_a}\bra{g_a} + E_{e_a} \ket{e_a}\bra{e_a}.
\end{align}
The coupling between the two donors is modeled as a Heisenberg XXZ spin chain augmented by a DM term. In this model, the exchange interactions along the \(x\) and \(y\) directions are characterized by the same strength \(J\), while the interaction along the \(z\) direction is controlled by the anisotropy parameter \(\kappa\). The XXZ Hamiltonian is given by~\cite{r53}:
\begin{equation}
H_{\text{XXZ}} = J \left( \sigma_x \otimes S_x + \sigma_y \otimes S_y + \kappa \, \sigma_z \otimes S_z \right),
\end{equation}
wherein the operators \(\sigma_i\) denote the standard Pauli matrices for the qubit, given by
\begin{align}
\sigma_x = \begin{pmatrix} 0 & 1 \\ 1 & 0 \end{pmatrix}, \quad
\sigma_y = \begin{pmatrix} 0 & -i \\ i & 0 \end{pmatrix}, \quad
\sigma_z = \begin{pmatrix} 1 & 0 \\ 0 & -1 \end{pmatrix},
\end{align}
and \(S_i\) are the spin-1 operators for the qutrit, explicitly expressed as
\begin{equation}
S_x = \frac{1}{\sqrt{2}} \begin{pmatrix} 0 & 1 & 0 \\ 1 & 0 & 1 \\ 0 & 1 & 0 \end{pmatrix}, \;
S_y = \frac{1}{\sqrt{2}} \begin{pmatrix} 0 & -i & 0 \\ i & 0 & -i \\ 0 & i & 0 \end{pmatrix}, \;
S_z = \begin{pmatrix} 1 & 0 & 0 \\ 0 & 0 & 0 \\ 0 & 0 & -1 \end{pmatrix}.
\end{equation}

The sign of \(J\) determines the nature of the exchange: \(J > 0\) corresponds to antiferromagnetic coupling, while \(J < 0\) account for ferromagnetic coupling. The spin-1 operators \(S^i\) serve as a convenient matrix basis for the three-level donor. They do not represent real spin sublevels but an effective three-level electronic system, with the mapping \(|{-1}\rangle \leftrightarrow |g_D\rangle\), \(|0\rangle \leftrightarrow |e_D\rangle\), and \(|{+1}\rangle \leftrightarrow |u_D\rangle\). In this representation, the exchange and DM terms describe effective excitonic coupling between the qubit and the qutrit, with \(S^{\pm}\) connecting levels that are adjacent in this ordering.
In addition to the exchange interaction, the Hamiltonian includes a DM term, which arises from inversion symmetry breaking in magnetic systems. The DM interaction tends to align spins perpendicular to each other and is expressed as~\cite{r53}:
\begin{equation}
H_{\text{DM}} = D_z \left( \sigma_x \otimes S_y - \sigma_y \otimes S_x \right),
\end{equation}
where \(D_z\) denotes the strength of the DM interaction along the \(z\) direction. This term not only enriches the phase behavior of the system but also provides a practical handle for tuning the output photocurrent.

The total Hamiltonian of the system is thus the sum of these three contributions:
\begin{equation}
H = H_{\text{self}} + H_{\text{XXZ}} + H_{\text{DM}}.
\end{equation}
This Hamiltonian constitutes the core of the dynamical model. The initial state of the complete donor-acceptor system is chosen as the ground state of the total Hamiltonian. To construct this state, we first diagonalize the donors Hamiltonian \(H_{\text{donors}} = H_d + H_D + H_{\text{XXZ}} + H_{\text{DM}}\) to find its ground state \(|\psi_{\text{donors}}\rangle\).The acceptor is placed in its ground state \(|g_a\rangle\). The initial state of the full system is then given by the tensor product:
\begin{equation}
|\psi_0\rangle = |\psi_{\text{donors}}\rangle \otimes |g_a\rangle, \qquad \rho_0 = |\psi_0\rangle\langle\psi_0|.
\label{eq:initial_state}
\end{equation}
This represents a natural physical initial condition, corresponding to a system that has reached equilibrium before the optical excitation is turned on.

In the following section, the coupling of the system to photonic and phononic reservoirs is incorporated and the Lindblad master equation governing its open-system dynamics is derived.
Being in contact with both photonic and phononic reservoirs, the quantum photocell introduced above is an open quantum system that continuously exchanges energy and particles with its environment. Consequently, its dynamics cannot be adequately described by the Schr\"odinger equation alone, and a master-equation framework for open quantum systems is required~\cite{r54}. In this work, we employ the Lindblad master equation, which governs the time evolution of the system density matrix \(\rho\) as follows:
\begin{equation}
\dot{\rho} = -i[H, \rho] + \mathcal{D}[\rho],
\label{eq:lindblad}
\end{equation}
where the first term, \(-i[H, \rho]\), accounts for the unitary evolution generated by the total Hamiltonian \(H\). The second term, \(\mathcal{D}[\rho]\), known as the dissipator, models the effects of decoherence and dissipation arising from the system-environment interaction. It is explicitly given by
\begin{equation}
\mathcal{D}[\rho] = \sum_k \left( 2 C_k \rho C_k^\dagger - C_k^\dagger C_k \rho - \rho C_k^\dagger C_k \right),
\label{eq:dissipator}
\end{equation}
where the operators \(C_k\) (\(C_k^\dagger\)) are the lowering (raising) operators, each corresponding to a distinct physical process occurring in the system. Properly identifying these operators is essential for accurately modeling the system dynamics.

It is important to emphasize that the photonic reservoir represents continuous and stationary illumination, rather than a pulsed excitation. However, continuous illumination does not imply that the system is immediately in its steady state. At \(t=0\), the donors and acceptor have their prescribed initial populations (the ground state). The photonic reservoir then excites the donors, the excitations transfer to the acceptor, and the external load removes them. These competing processes take time to establish a steady flow. Although the photocell is continuously illuminated, its initial state is not generally the non-equilibrium steady state. The transient photocurrent reflects the redistribution of populations and the establishment of excitation transfer and extraction processes following the initial preparation. The steady-state current is reached when these processes balance.

To model the interaction with the reservoirs, the Bose-Einstein distribution is employed to compute the mean occupation number of the reservoir mode resonant with the corresponding system transition~\cite{r23}. This quantity, denoted by \(\bar{n}\), is defined as
\begin{equation}
\bar{n}(\omega, T) = \frac{1}{\exp\left(\frac{\hbar \omega}{k_B T}\right) - 1},
\label{eq:planck}
\end{equation}
where \(\omega\) is the transition frequency and \(T\) is the reservoir temperature. For each optical and phononic transition, \(\bar{n}\) is evaluated using the corresponding energy gap and the appropriate reservoir temperature—\(T_{\text{ph}}\) for photons and \(T_{\text{pn}}\) for phonons. These thermal occupation numbers determine the absorption and stimulated-emission rates appearing in the Lindblad collapse operators and thereby govern the dissipative dynamics of the system.

The Lindblad master equation is solved numerically using QuTiP~\cite{r57}, which employs an adaptive integration scheme. The simulations are performed over the time interval \(t=0\) to \(t=15\) (in natural units, \(\hbar=1\)) using 500 time points. The number of integration steps is set to \(10^5\), ensuring convergence. We have verified that both the peak transient current and the steady-state current are insensitive to further reductions in the time step or extensions of the time window.

In the following, the collapse operators corresponding to each relevant physical process are introduced.

\subsection{Photonic Processes in the First Donor (Qubit)}

The first donor, modeled as a qubit, interacts with the photonic reservoir, enabling both photon emission and absorption. The corresponding collapse operators are defined separately for absorption and emission as
\begin{align}
C_{1}^{\text{abs}} &= \sqrt{\gamma_d n_d} \, \sigma_+^{(d)}, &
C_{1}^{\text{em}} &= \sqrt{\gamma_d (n_d + 1)} \, \sigma_-^{(d)},
\end{align}
where \(\sigma_+^{(d)} = \ket{e_d}\bra{g_d}\) and \(\sigma_-^{(d)} = \ket{g_d}\bra{e_d}\). These processes correspond to the orange double-headed arrows in the schematic diagram of Figure~\ref{f2}.

\subsection{Photonic Processes in the Second Donor (Qutrit)}

The second donor, a qutrit, supports two distinct optical transitions: one between \(\ket{g_D}\) and \(\ket{e_D}\), and another between \(\ket{g_D}\) and \(\ket{u_D}\). For each transition, absorption and emission are treated as independent collapse channels:
\begin{align}
C_{2}^{\text{abs}} &= \sqrt{\gamma_{ge} n_{ge}} \, L_{eg}^\dagger, &
C_{2}^{\text{em}} &= \sqrt{\gamma_{ge} (n_{ge} + 1)} \, L_{eg}, \\
C_{3}^{\text{abs}} &= \sqrt{\gamma_{gu} n_{gu}} \, L_{ug}^\dagger, &
C_{3}^{\text{em}} &= \sqrt{\gamma_{gu} (n_{gu} + 1)} \, L_{ug},
\end{align}
where \(L_{eg} = \ket{g_D}\bra{e_D}\) and \(L_{ug} = \ket{g_D}\bra{u_D}\). Here, \(\gamma_{ge}\) and \(\gamma_{gu}\) denote the respective decay rates, while \(n_{ge}\) and \(n_{gu}\) represent the mean photon numbers at the corresponding transition frequencies. These transitions are represented by the orange double-headed arrows in Figure~\ref{f2}.

\subsection{Interference Terms in the Second Donor}

When the two optical transitions are coupled to a common bath and possess non-orthogonal transition dipole moments, quantum interference between the transition pathways gives rise to cross-coupling terms in the master equation. The emission cross terms are given by
\begin{align}
C_{4}^{\text{em},1} &= \frac{\gamma_x}{2} \left( 2\sigma_{ge}\rho\sigma_{ug} - \sigma_{ue}\rho - \rho\sigma_{ue} \right), \label{eq:em1} \\
C_{4}^{\text{em},2} &= \frac{\gamma_x}{2} \left( 2\sigma_{gu}\rho\sigma_{eg} - \sigma_{eu}\rho - \rho\sigma_{eu} \right), \label{eq:em2}
\end{align}
while the absorption cross terms read
\begin{align}
C_{4}^{\text{ab},1} &= \frac{\gamma_y}{2} \left( 2\sigma_{ug}\rho\sigma_{ge} \right), \label{eq:ab1} \\
C_{4}^{\text{ab},2} &= \frac{\gamma_y}{2} \left( 2\sigma_{eg}\rho\sigma_{gu} \right), \label{eq:ab2}
\end{align}
where \(\sigma_{ij} = \ket{i}\bra{j}\) are transition operators between the qutrit levels. The interference coupling rates \(\gamma_x\) and \(\gamma_y\) are defined as
\begin{equation}
\gamma_x = \sqrt{\gamma_{ge} \gamma_{gu} (n_{ge} + 1)(n_{gu} + 1)}, \qquad
\gamma_y = \sqrt{\gamma_{ge} \gamma_{gu} n_{ge} n_{gu}}.
\end{equation}
These terms capture reservoir-induced interference between the two optical transitions. In Figure~\ref{f2}, these rates are collectively denoted by \(\gamma_c\) and are represented by orange dashed double-headed arrows.

The interference cross-terms in Eqs.~\eqref{eq:em1}--\eqref{eq:ab2} depend on the relative orientation of the transition dipole moments of the two optical transitions. Following the standard treatment of dipole-dipole interference in three-level systems~\cite{r54}, the cross-damping rate is given by
\begin{equation}
\gamma_{eu} = \frac{2\sqrt{\omega_{ge}^3\omega_{gu}^3}}{3\hbar c^3} \, \vec{\mu}_{ge} \cdot \vec{\mu}_{gu} = \beta_I \sqrt{\gamma_{ge}\gamma_{gu}},
\end{equation}
where \(\vec{\mu}_{ge}\) and \(\vec{\mu}_{gu}\) are the transition dipole moments, and
\begin{equation}
\beta_I = \frac{\vec{\mu}_{ge} \cdot \vec{\mu}_{gu}}{|\vec{\mu}_{ge}| |\vec{\mu}_{gu}|} = \cos\theta,
\end{equation}
with \(\theta\) being the angle between the two transition dipole moments. In our model, we assume that the dipoles are parallel (\(\theta = 0\), \(\beta_I = 1\)), which corresponds to the maximum interference between the two transition pathways. This choice is consistent with the form of the interference rates in Eqs.~\eqref{eq:em1}--\eqref{eq:ab2} and is a standard assumption in quantum photocell models.\\

\subsection{Phonon-Assisted Transfers from Donors to Acceptor}

Electron transfer from each donor to the acceptor is mediated by interactions with a phononic bath. These processes are modeled via three independent transfer channels, each with separate emission and absorption operators:
\begin{align}
C_{6}^{\text{em}} &= \sqrt{\gamma_{f_1} (n_{f_1} + 1)} \, L_{f_1}, &
C_{6}^{\text{abs}} &= \sqrt{\gamma_{f_1} n_{f_1}} \, L_{f_1}^\dagger, \\
C_{7}^{\text{em}} &= \sqrt{\gamma_{f_2} (n_{f_2} + 1)} \, L_{f_2}, &
C_{7}^{\text{abs}} &= \sqrt{\gamma_{f_2} n_{f_2}} \, L_{f_2}^\dagger, \\
C_{8}^{\text{em}} &= \sqrt{\gamma_{f_3} (n_{f_3} + 1)} \, L_{f_3}, &
C_{8}^{\text{abs}} &= \sqrt{\gamma_{f_3} n_{f_3}} \, L_{f_3}^\dagger,
\end{align}
with \(L_{f_1} = \ket{e_a}\bra{e_d}\), \(L_{f_2} = \ket{e_a}\bra{e_D}\), and \(L_{f_3} = \ket{e_a}\bra{u_D}\). The parameters \(\gamma_{f_1}, \gamma_{f_2}, \gamma_{f_3}\) denote the phonon-assisted transfer rates, and \(n_{f_1}, n_{f_2}, n_{f_3}\) are the corresponding mean phonon occupation numbers. The factors \((n_{f_i}+1)\) and \(n_{f_i}\) account for phonon emission and absorption, respectively. These processes appear as blue double-headed arrows in Figure~\ref{f2}.

The transfer rates in Eqs.~(25)--(27) are phenomenological and follow the standard Born--Markov, weak-coupling (Fermi golden-rule) treatment of a system coupled to a thermal phonon bath, as commonly used in quantum photocell models~\cite{r23,r24,r50}. The downward and upward rates scale as $\gamma_{fi}(\bar{n}_{fi}+1)$ and $\gamma_{fi}\bar{n}_{fi}$, respectively, which satisfies detailed balance relation. Nonadiabatic Marcus theory and its extension, the Marcus--Levich--Jortner theory, which adds quantum high-frequency vibrational modes~\cite{r55,r56}, describe the strong-coupling regime, where the rate is set by the electronic coupling, the driving force, and the reorganization energy. In both cases, these quantities combine into a single effective rate per donor--acceptor pair, and $\gamma_{f1}$, $\gamma_{f2}$, and $\gamma_{f3}$ play this role in our model. The donor--acceptor gaps ($0.1$, $0.2$, and $0.3\,\mathrm{eV}$), as expressed in Table I, are several times larger than $k_B T_{\mathrm{pn}} \approx 0.026\,\mathrm{eV}$ ($300\,\mathrm{K}$), so back-transfer is suppressed and the transfer is effectively downward. Our results should therefore be read as qualitative trends, and quantitative predictions would require rates for a specific material. The robustness of these trends against variations in $\gamma_f$ and $T_{\mathrm{pn}}$ is tested in Section V.

\subsection{Decay to External Load and Photocurrent Generation}

Once the electron reaches the acceptor, it relaxes to the lower energy level through an external load, thereby generating photocurrent. This process is described by the collapse operator
\begin{equation}
C_{9} = \sqrt{\Gamma} \, \sigma_a, \qquad \sigma_a = \ket{g_a}\bra{e_a},
\label{eq:c9}
\end{equation}
where $\Gamma$ is the decay rate to the load, which directly controls the output photocurrent. This transition is indicated by the green arrow in the schematic diagram of Figure~\ref{f2}.

\subsection{Recombination (Loss) Processes}

Two types of recombination processes are considered: direct recombination from the acceptor to the second donor, and ground-state recombination. These are modeled as
\begin{align}
C_{10} &= \sqrt{\gamma_{r_1}} \, L_{r_1}, & L_{r_1} &= \ket{g_D}\bra{e_a}, \\
C_{11}^{\text{em}} &= \sqrt{\gamma_{r_2} (n_{r_2} + 1)} \, L_{r_2}, &
C_{11}^{\text{abs}} &= \sqrt{\gamma_{r_2} n_{r_2}} \, L_{r_2}^\dagger,
\end{align}
where $L_{r_2} = \ket{g_D}\bra{g_a}$. Here, $\gamma_{r_1}$ and $\gamma_{r_2}$ are the recombination rates, and $n_{r_2}$ is the mean phonon occupation at the corresponding transition energy. In Figure~\ref{f2}, direct recombination ($\gamma_{r_1}$) is shown as a unidirectional blue arrow, while ground-state recombination ($\gamma_{r_2}$) is represented by a blue double-headed arrow, consistent with its phonon-assisted nature.

With this complete set of collapse operators, the Lindblad master equation is fully specified. Its numerical solution yields the time evolution of the density matrix, from which key observables—photocurrent, voltage, output power, and purity—can be evaluated.

\section{\label{sec:level5}RESULTS AND DISCUSSION}
In this section, the numerical results obtained from solving the Lindblad master equation are presented. The parameters used in the simulations are listed in Table~\ref{tab:params}. In our numerical simulations, natural units are employed, such that \(k_B = \hbar = 1\). In these units, all energies and temperatures are expressed in electron-volts (eV), and all decay rates (\(\gamma\) and \(\Gamma\)) are in s\(^{-1}\). The unit of time is defined as \(1/\Gamma\), which, with \(\Gamma = 1\) eV, corresponds to \(1/\Gamma = \hbar/\text{eV} \approx 0.658\) fs.

\begin{figure}[!ht]
\centering
\includegraphics[width=1\columnwidth]{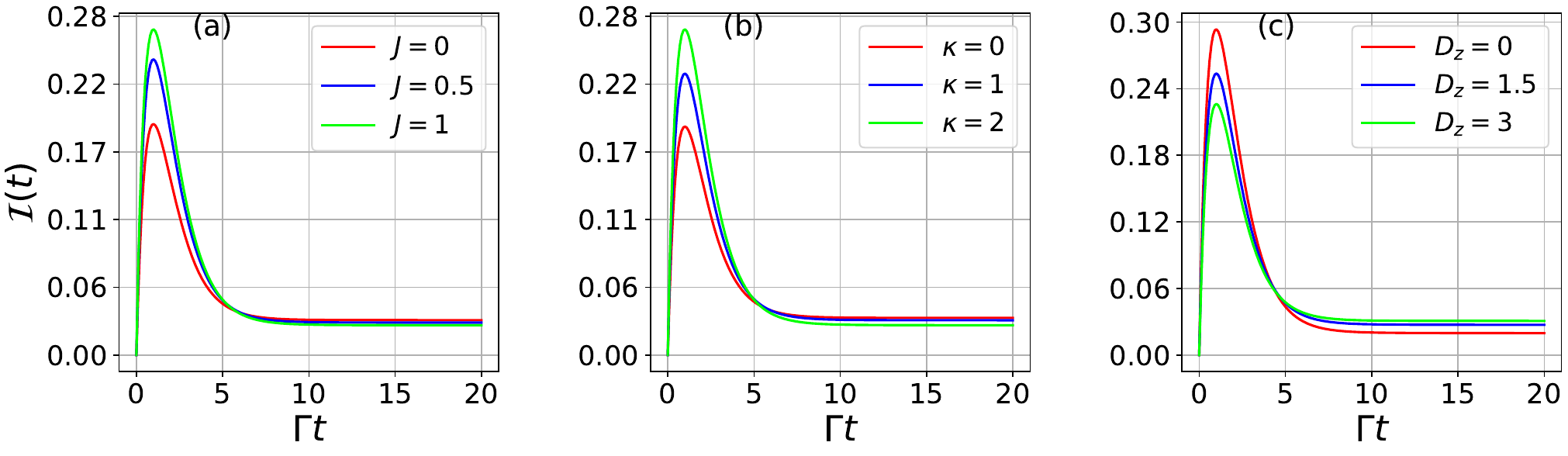}
\caption{Time evolution of the photocurrent $\mathcal{I}(t)$ for different values of the coupling parameters: (a) dependence on $J$ with $\kappa=2$ and $D_z=1$; (b) dependence on $\kappa$ with $J=1$ and $D_z=1$; (c) dependence on $D_z$ with $J=1$ and $\kappa=2$. Time is in units of \(1/\Gamma\) with \(\Gamma = 1\) eV, so \(1/\Gamma \approx 0.658\) fs.}
\label{f3}
\end{figure}

Therefore, all time values in the time-dependent figures (Figures~\ref{f3}, \ref{f7}, \ref{f8}, and \ref{f9}) are expressed in this unit. The goal is to examine how the coupling parameters affect the dynamical behavior of the photocell and to clarify the distinct contribution of each interaction channel to the photocurrent.

\begin{table}[htbp]
\centering
\scalebox{1.3}{%
\begin{tabular}{|c|c|}
\hline
\textbf{Parameter} & \textbf{Value} \\
\hline
$E_{g_d}$ & 0.3 \\
$E_{e_d}$ & 1.6 \\
$E_{g_D}$ & 0.3 \\
$E_{e_D}$ & 1.7 \\
$E_{u_D}$ & 1.8 \\
$E_{g_a}$ & 0.4 \\
$E_{e_a}$ & 1.5 \\
\hline
$\gamma_d$ & 0.01 \\
$\gamma_{ge}$ & 1.0 \\
$\gamma_{gu}$ & 0.01 \\
$\gamma_{f_1}$ & 1.0 \\
$\gamma_{f_2}$ & 0.01 \\
$\gamma_{f_3}$ & 0.01 \\
$\gamma_{r_1}$ & 0.01 \\
$\gamma_{r_2}$ & 0.01 \\
\hline
$k_B T_{\text{ph}}$ & 0.517 \\
$k_B T_{\text{pn}}$ & 0.02585\\
\hline
\end{tabular}%
}
\caption{Parameters used in the numerical simulations. All energies are in units of eV, and all rates are in units of s$^{-1}$. Temperatures are given as \(k_B T\) in eV (with \(k_B = 1\)).}
\label{tab:params}
\end{table}
The time evolution of the photocurrent for different values of the exchange coupling \(J\), the anisotropy \(\kappa\), and the DM interaction \(D_z\) is shown in Figure~\ref{f3}, for panels (a), (b), and (c), respectively. As shown in Figure~\ref{f3}(a), increasing the exchange coupling \(J\) between the two donors significantly enhances the generated photocurrent. 

In this sense, \(J\) acts as a quantum bridge that expedites charge transport. A similar, though milder, trend is observed in Figure~\ref{f3}(b) for the anisotropy parameter \(\kappa\). By controlling the relative weight of the \(z\)-direction exchange, increasing \(\kappa\) drives the system toward a more Ising-like regime; consequently, the excited-state lifetime is extended, and the photocurrent rises. In contrast, Figure~\ref{f3}(c) reveals a markedly different behavior for the DM interaction \(D_z\). Unlike \(J\) and \(\kappa\), increasing \(D_z\) suppresses the photocurrent. The highest current is observed at \(D_z=0\), where the system is governed solely by the exchange interaction. Although \(D_z\) is detrimental in the dynamical regime, we will later show that its role reverses in the steady state, where it can enhance the current.
\begin{figure}[!ht] 
\centering
\includegraphics[width=1.02\columnwidth]{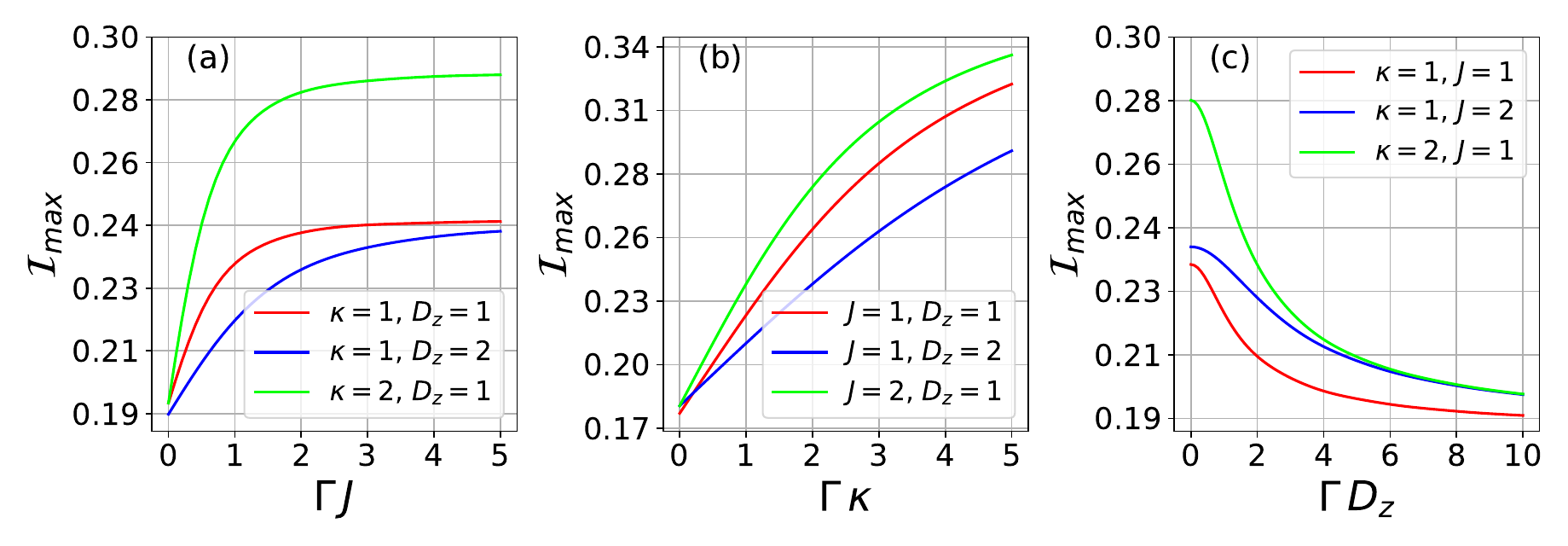}
\caption{Maximum temporal photocurrent $\mathcal{I}_{\max}$ as a function of the coupling parameters: (a) dependence on $J$, (b) dependence on $\kappa$, and (c) dependence on $D_z$. Each panel is plotted for fixed values of the other parameters to highlight the role of each interaction in determining the peak current accessible during the time evolution.}
\label{f4}
\end{figure}

To gain a clearer understanding of the role played by the coupling parameters, Figure~\ref{f4} displays the maximum temporal photocurrent as a function of \(J\), \(\kappa\), and \(D_z\). Figures~\ref{f4}(a) and (b) show that increasing \(J\) and \(\kappa\), particularly when the detrimental effect of \(D_z\) is suppressed, facilitates phonon-assisted transfer and thereby enhances the peak current. In contrast, panel Figure~\ref{f4}(c) reveals that increasing \(D_z\) degrades the current by disrupting the coherent spin order. However, at lower values of \(D_z\), the compensating role of stronger anisotropy becomes evident, as increasing \(\kappa\) can partially offset the suppressing effect of DM. 

\begin{figure}[htbp]
\centering
\includegraphics[width=0.5\textwidth]{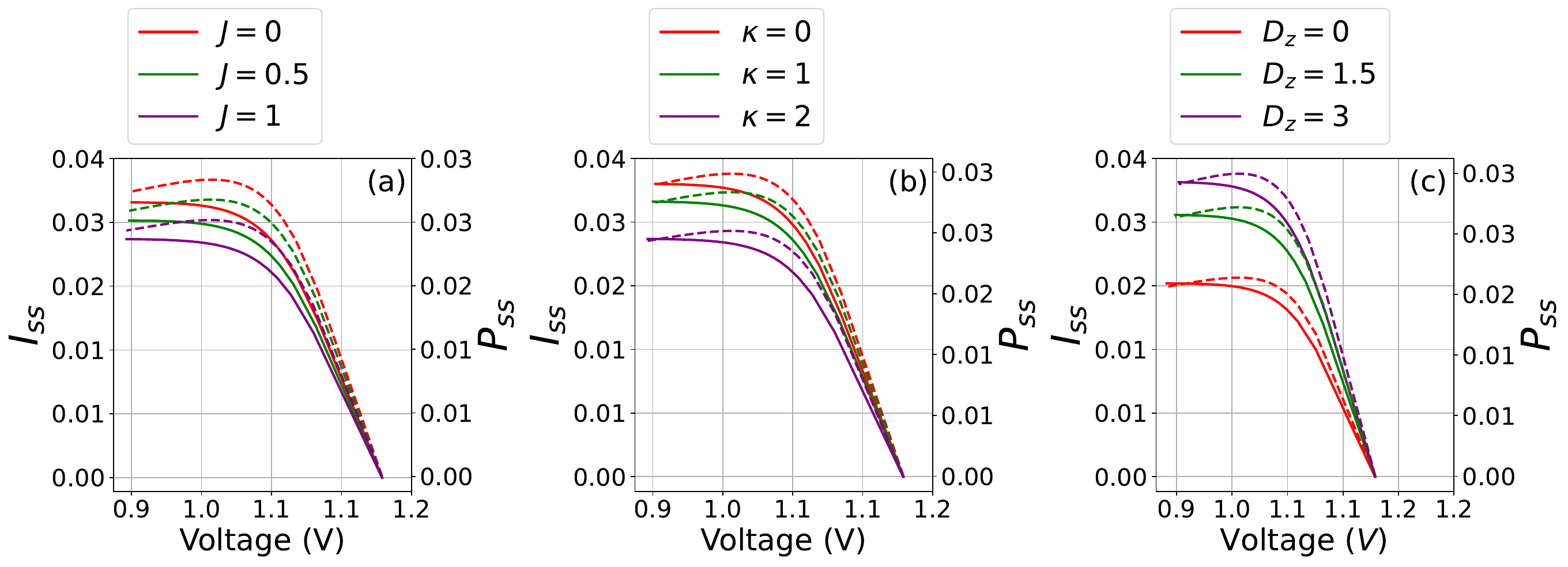}
\caption{Steady-state current and power curves ($I_{ss}$ and $P_{ss}$) as a function of voltage ($V$), for different values of the parameters $J$, $\kappa$, and $D_z$. In each panel, solid lines represent the current and dashed lines represent the output power. These curves are obtained by varying the decay rate to the external load ($\Gamma$).}
\label{f5}
\end{figure}

In a photovoltaic cell, the final usable output in the external circuit is determined in the steady-state regime by varying the decay rate to the external load (\(\Gamma\)). Changing \(\Gamma\) modifies the output voltage and current, yielding the current–voltage (\(I\)–\(V\)) characteristic curve in the steady state—one of the most important figures of merit for evaluating light-harvesting systems performance. The output power (\(P = I \cdot V\)) is also assessed in the same regime. In Figure~\ref{f5}, these results are presented for different values of the coupling parameters \(J\), \(\kappa\), and \(D_z\) in order to investigate the role of these interactions in the steady-state behavior of the system. 

As seen in Figures~\ref{f5}(a) and (b), in contrast to the transient regime, increasing \(J\) and \(\kappa\) in the steady state substantially reduces the current. In contrast, Figure~\ref{f5}(c) reveals a completely opposite behavior for the DM interaction, such that increasing \(D_z\) in the steady state enhances both the current and the output power. The output power (\(P_{ss}\)), following directly from \(P = I \cdot V\), exhibits a trend similar to that of the current. These results demonstrate that by properly selecting the operational regime and simultaneously tuning the coupling parameters, the system can be steered toward markedly different—and even contrasting—performance outcomes.

\begin{figure}[htbp]
\centering
\includegraphics[width=0.5\textwidth, height=0.25\textheight]{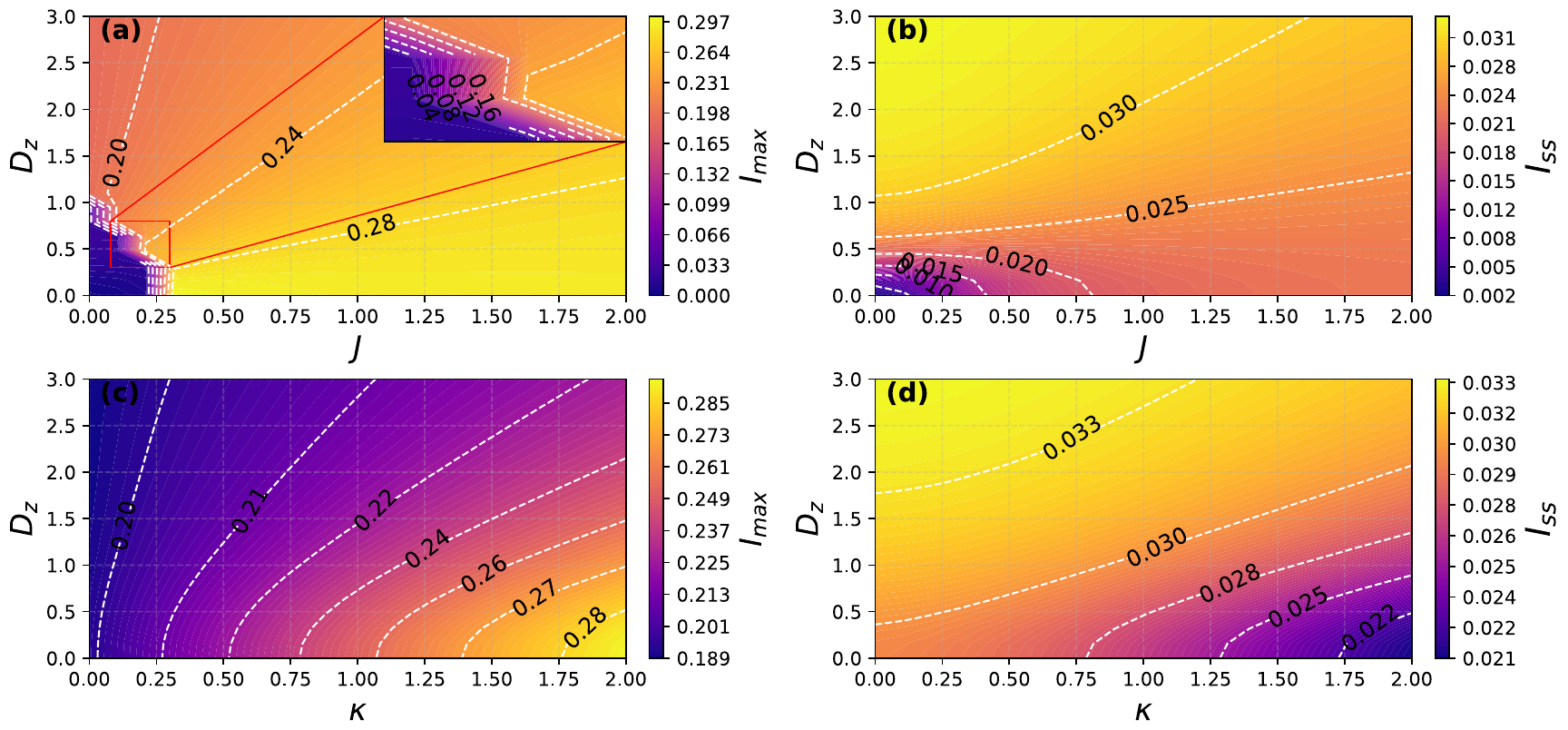}
\caption{Contour plots of the maximum (temporal) and steady-state photocurrents ($I_{ss}$) as functions of the coupling parameters. (a) and (c) show the maximum current ($I_{max}$) versus $J$ and $D_z$, and versus $\kappa$ and $D_z$, respectively. (b) and (d) display the corresponding steady-state currents for the same parameter combinations. White contour lines indicate constant current values.}
\label{f6}
\end{figure}

\begin{figure}[htbp]
\centering
\includegraphics[width=0.47\textwidth, height=0.2\textheight]{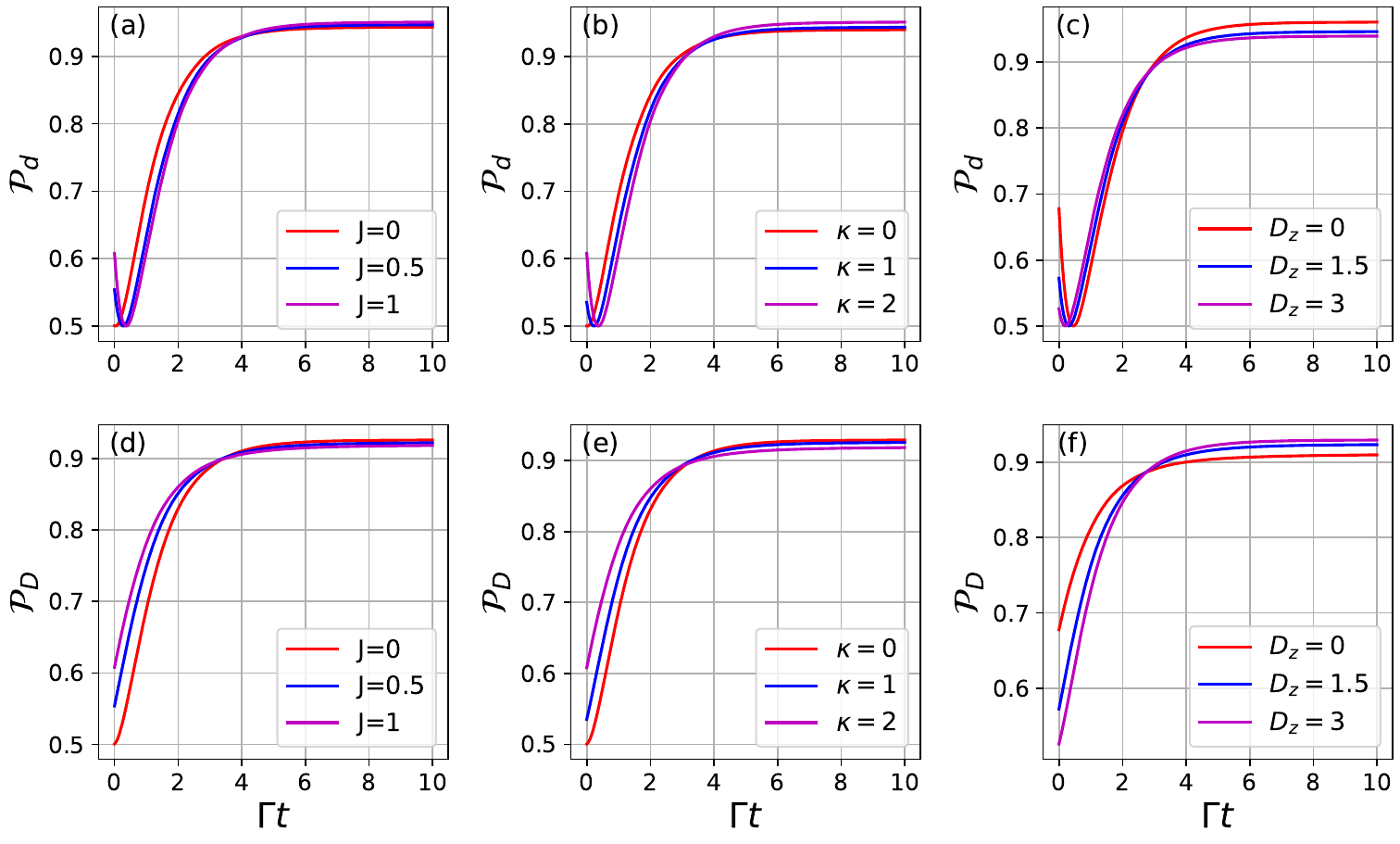}
\caption{Time dependence of the purity of the donor subsystems for different values of the coupling parameters $J$, $\kappa$, and $D_z$. (a)--(c) show the purity of the first donor (qubit), while (d)--(f) present the purity of the second donor (qutrit), corresponding to different values of $J$, $\kappa$, and $D_z$, respectively. Time is in units of \(1/\Gamma\) with \(\Gamma = 1\) eV, so \(1/\Gamma \approx 0.658\) fs.}
\label{f7}
\end{figure}

\begin{figure}[!ht]
\centering
\includegraphics[width=0.5\textwidth]{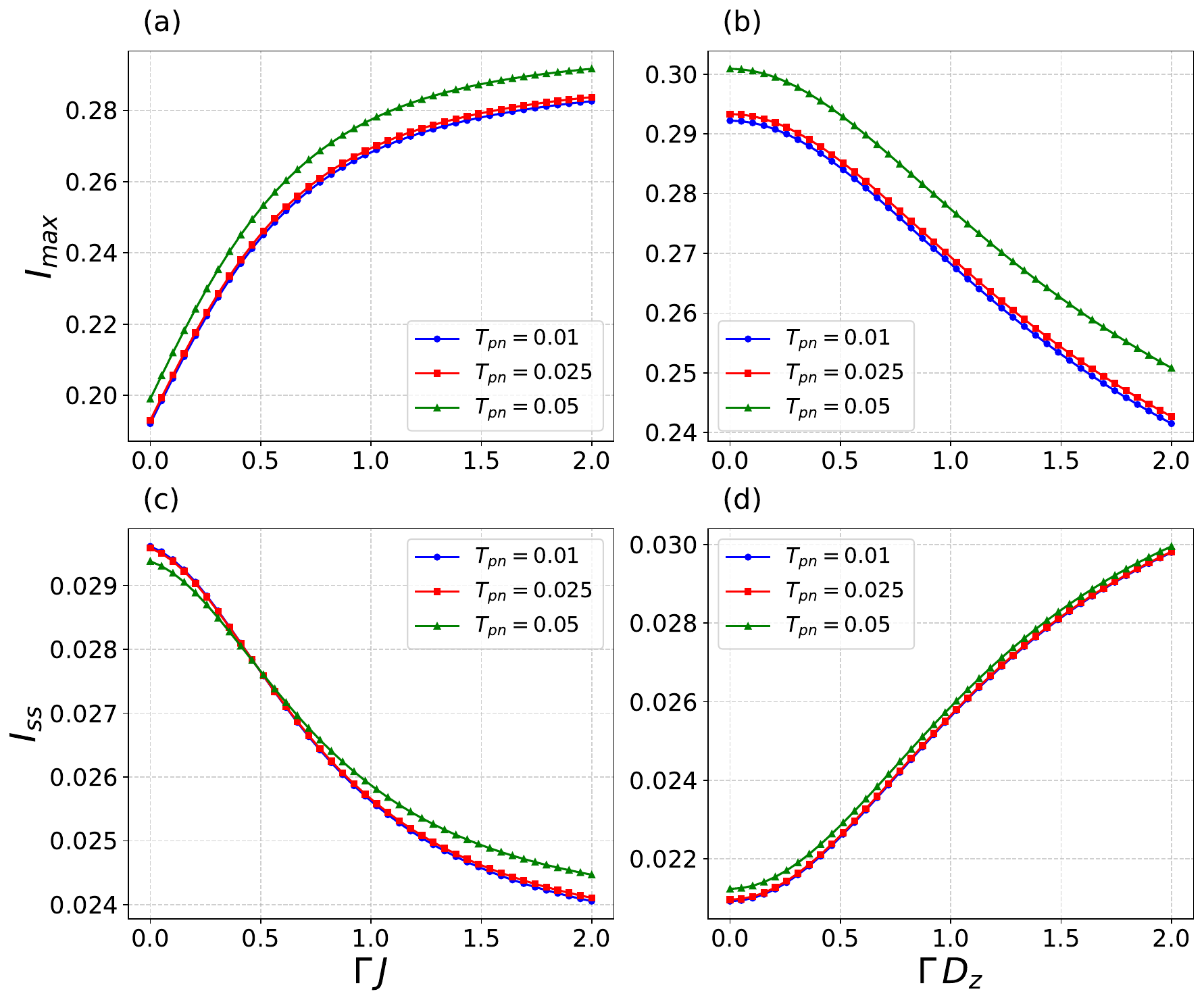}
\caption{Sensitivity of the photocurrent to the phonon temperature. (a) Maximum transient current (\(I_{max}\)) versus \(J\) (with \(\kappa=2\) and \(D_z=1\)) and (b) versus \(D_z\) (with \(\kappa=2\) and \(J=1\)). (c) Steady-state current (\(I_{ss}\)) versus \(J\) (with \(\kappa=2\) and \(D_z=1\)) and (d) versus \(D_z\) (with \(\kappa=2\) and \(J=1\)). All panels are plotted for three different phonon temperatures (\(T_{pn} = 0.01, 0.025, 0.05\) eV). The dichotomy between the transient and steady-state regimes persists across all temperatures, and higher temperatures lead to higher overall currents.}
\label{f8}
\end{figure}

\begin{figure}[!ht]
\centering
\includegraphics[width=0.5\textwidth]{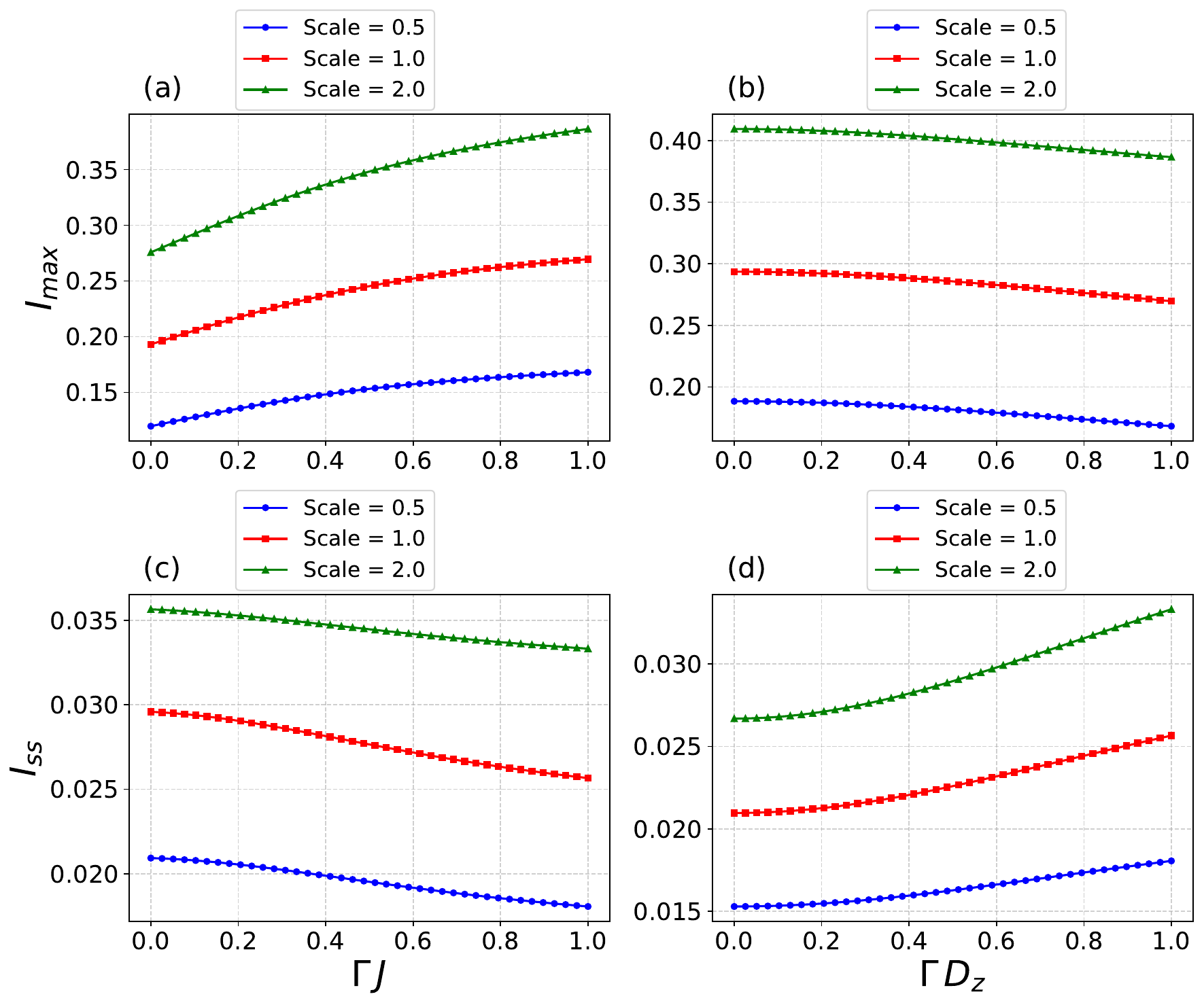}
\caption{Sensitivity of the photocurrent to the phonon-assisted transfer rates. (a) Maximum transient current (\(I_{max}\)) versus \(J\) (with \(\kappa=2\) and \(D_z=1\)) and (b) versus \(D_z\) (with \(\kappa=2\) and \(J=1\)). (c) Steady-state current (\(I_{ss}\)) versus \(J\) (with \(\kappa=2\) and \(D_z=1\)) and (d) versus \(D_z\) (with \(\kappa=2\) and \(J=1\)). All panels are plotted for three different scaling factors of the transfer rates ($\gamma_f$ scaled by 0.5, 1.0, and 2.0). The qualitative trends remain robust across all scaling factors, and larger transfer rates lead to higher overall currents.}
\label{f9}
\end{figure}

To enable a direct comparison of the photocurrent behavior in the transient and steady-state regimes, Figure~\ref{f6} presents contour plots of the maximum and steady-state currents as functions of the coupling parameters. Panels (a) and (c) show the maximum current versus \(J\) and \(D_z\), and versus \(\kappa\) and \(D_z\), respectively, while panels (b) and (d) display the corresponding steady-state currents for the same parameter combinations.

As shown in Figure~\ref{f6}(a), the maximum transient current increases monotonically with the exchange coupling \(J\). In contrast, the influence of the Dzyaloshinskii–Moriya interaction \(D_z\) depends strongly on the exchange-coupling regime. For moderate to large values of \(J\), increasing \(D_z\) suppresses the maximum transient current, whereas for weak exchange coupling the opposite trend is observed, with \(D_z\) leading to a slight enhancement of the current. Consequently, the highest transient current is obtained in the regime of strong exchange coupling and weak DM interaction, while the lowest current occurs when both \(J\) and \(D_z\) are small. In contrast to the transient regime, for the steady state, the influence of the exchange coupling depends on the strength of the DM interaction, as shown by Figure~\ref{f6}(b). For weak DM interaction, increasing \(J\) slightly enhances the steady-state current, whereas at larger values of \(D_z\), further increasing \(J\) suppresses the stationary current. By comparison, the DM interaction exhibits a monotonic influence throughout the parameter space, with the steady-state current increasing as \(D_z\) increases. Consequently, the maximum steady-state current is achieved for weak exchange coupling and strong DM interaction, whereas the minimum current occurs when both \(J\) and \(D_z\) are small.

As depicted in Figure~\ref{f6}(c), increasing the donor-donor coupling strength \(\kappa\) substantially enhances the maximum transient current, particularly in the weak-DM regime. The largest transient current is therefore obtained for strong anisotropy and small values of \(D_z\). However, this enhancement is gradually suppressed as the DM interaction increases, leading to a continuous decrease in the transient current over the entire parameter range. In contrast, the steady-state current shown in Figure~\ref{f6}(d) exhibits the reverse dependence on the system parameters. Increasing \(\kappa\) gradually suppresses the stationary current, whereas increasing the DM interaction significantly improves the long-time transport efficiency. Consequently, the highest steady-state current is obtained in the region characterized by strong DM interaction and relatively weak anisotropy in donors coupling. The opposite trends observed in Figure~\ref{f6}(c) and (d) further confirm that the optimal parameter regime for transient transport differs substantially from that required to maximize the steady-state performance of the quantum photocell.

To investigate the the influence of internal interactions on the local state of the donors, the time evolution of the purity(as defined in Equation~\eqref{eq4}) of the first and second donors is presented in Figure~\ref{f7} for different values of the interaction parameters $J$, $\kappa$, and $D_z$. As observed, the purity of the two donors exhibits distinct behaviors in the transient and steady-state regimes. For the first donor (qubit), increasing $J$ and $\kappa$ reduces the purity at early times, while increasing $D_z$ enhances it. However, upon reaching the steady state, this trend reverses completely: larger $J$ and $\kappa$ are accompanied by increased purity, whereas larger $D_z$ leads to a decrease. In contrast, the purity of the second donor (qutrit) displays a more regular behavior, such that its variations in the transient regime follow the trend of the transient current, and in the steady state, they align with the steady-state current behavior. 
Unlike the qutrit, the qubit purity does not exhibit a consistent correlation with the photocurrent across both regimes; its response changes sign depending on the temporal regime. We stress that purity is a descriptive indicator of mixedness: a change in the local purity of a donor can arise from classical redistribution of populations, from coherences between its levels, or from correlations (including entanglement) with the other subsystem, and purity alone cannot distinguish these contributions. The correlation between the qutrit purity and the photocurrent should therefore be read as an empirical observation and not as direct evidence of a quantum enhancement mechanism. The larger Hilbert space of the qutrit may contribute to this behavior, but establishing its role would require a dedicated analysis, which we leave for future work.

To assess the robustness of our predicted trends against variations in the environmental conditions, a comprehensive sensitivity analysis has performed by varying the phonon temperature and the phonon-assisted transfer rates. Figure~\ref{f8} shows the dependence of the maximum transient current (\(I_{max}\)) and the steady-state current (\(I_{ss}\)) on the interaction parameters \(J\) and \(D_z\) for three different phonon temperatures (\(T_{pn} = 0.01, 0.02585, 0.05\) eV). In Figure~\ref{f9}, the same analysis is repeated for three different scaling factors of the phonon-assisted transfer rates (\(\gamma_{f1}\), \(\gamma_{f2}\), and \(\gamma_{f3}\)). To examine how the overall strength of the phonon-assisted transfer channel affects the system dynamics, we scale all three transfer rates simultaneously by a common factor. Specifically, the base values of the three rates (as given in Table~\ref{tab:params}) are multiplied by 0.5, 1.0, and 2.0. This means that:

\noindent For the scaling factor 0.5, all three rates are reduced to half of their original values;

\noindent For the scaling factor 1.0, the original values (as listed in Table~\ref{tab:params}) are used;

\noindent For the scaling factor 2.0, all three rates are doubled.

In all three cases, the relative ratios between the three rates ($\gamma_{f1} : \gamma_{f2} : \gamma_{f3}$) remain fixed, so that the only variable is the overall strength of the phonon-transfer channel.

The results demonstrate that the distinction between the transient and steady-state regimes is preserved independently of the environmental conditions. In all cases, increasing \(J\) enhances the transient current while suppressing the steady-state current, whereas \(D_z\) exhibits the opposite behavior. Furthermore, at any fixed value of the interaction parameters, increasing the temperature or the transfer rates leads to an increase in the current in both the transient and steady-state regimes. This behavior indicates that, within the parameter regime under consideration, the phonon-assisted transfer channel dominates over recombination losses, and the system operates as an efficient photocell. These observations confirm the robustness of our results and provide deeper physical insights into the interplay between spin interactions and environmental coupling.

\section{\label{sec:level6}Conclusion}
In this work, the performance of a hybrid quantum photocell composed of a qubit-qutrit donor and an acceptor has been investigated in the presence of Heisenberg exchange, anisotropic coupling, and Dzyaloshinskii–Moriya spin–orbit interaction between the donors. The results establish that the coupling parameters play a decisive role in controlling the photocell behavior in both transient and steady-state regimes. Increasing the exchange coupling \(J\) and the anisotropy \(\kappa\) enhances the transient current while suppressing the steady-state current, whereas the Dzyaloshinskii–Moriya interaction \(D_z\) exhibits a strictly opposite effect on these two temporal regimes. The purity analysis further shows that the qutrit’s local purity correlates consistently with the current variations in both the transient and steady-state regimes, whereas the qubit’s does not. This is a descriptive observation, since purity alone cannot separate population redistribution, coherence, and correlations with the other subsystem, and establishing the qutrit’s role would require a dedicated analysis.

The results presented here demonstrate that hybrid qubit–qutrit architectures, by offering an extended Hilbert space and enhanced tunability, provide a promising route for controlling energy-transfer dynamics and optimizing next-generation quantum photovoltaic devices. The discrete-level structure and the possibility of tuning the coupling parameters in these systems open up new opportunities for exploring energy storage and may serve as a first step toward investigating their potential as quantum batteries. These results also pave the way for future research on multilevel systems and spin-based interactions in the development of advanced quantum energy conversion and storage technologies.\\

\medskip
\noindent \textbf{Acknowledgements} \par
\noindent This work is based upon research funded by Iran National Science Foundation (INSF) under project No. 4042716.\\

\appendix
\section{Total Charge as a Complementary Figure of Merit}
\label{app:charge}

\begin{figure}[!ht]
\centering
  \includegraphics[width=1\columnwidth]{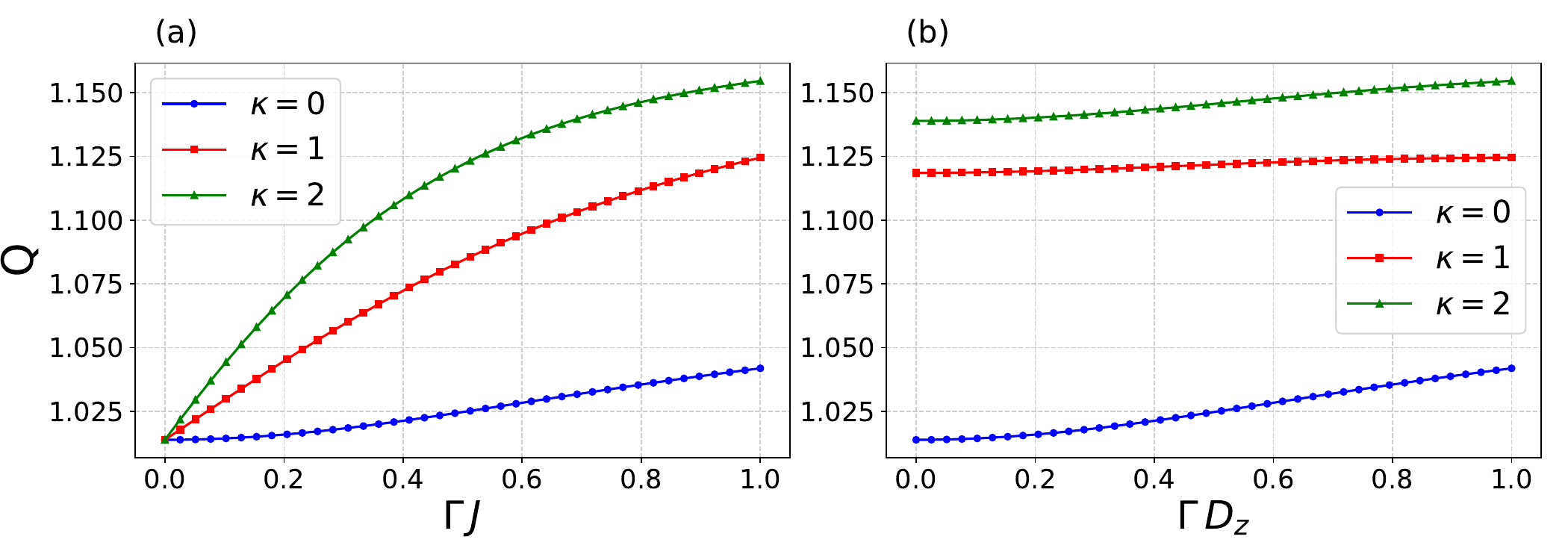}
  \caption{Total charge Q as a function of (a) \(J\) (with \(D_z=1\)) and (b) \(D_z\) (with \(J=1\)) for three different values of \(\kappa\) (0, 1, and 2).}
\label{f10}
\end{figure}

As shown in Figs.~3, increasing \(J\) and \(\kappa\) enhances the transient photocurrent, while \(D_z\) suppresses it. A larger peak current does not by itself imply a larger amount of extracted charge, so here we discuss when such a transient enhancement is useful in practice. It is relevant for pulsed or short-time operation, in which the cell is illuminated for a window \(T\) shorter than the time needed to reach the steady state, so that the output per cycle is set by the early-time dynamics. This is the situation in time-resolved and ultrafast excitation of coupled donor pairs (molecular dimers, chromophores, quantum dots), which is routinely realized on femtosecond-to-picosecond time scales~\cite{r58,r59}. In our units, time is measured in \(1/\Gamma=\hbar/\mathrm{eV}\approx0.658\)~fs. For continuous illumination (\(T\to\infty\)), the steady-state current governs the output instead. To complement the peak current, we therefore evaluate the total extracted charge over the observation window,
\begin{equation}
Q(T)=\int_0^{T} I(t)\,dt ,
\end{equation}
with \(T=20\), shown in Fig.~\ref{f10} as a function of \(J\) (left panel) and \(D_z\) (right panel) for \(\kappa=0,1,2\).
Figure~\ref{f10} shows the total charge \(Q\) as a function of \(J\) (left panel) and \(D_z\) (right panel) for three different values of \(\kappa\) (0, 1, and 2). For all three values of \(\kappa\), the total charge \(Q\) increases monotonically with \(J\). This indicates that a stronger exchange interaction extracts a larger total charge over time. The total charge also increases with \(D_z\) (with a gentler slope). This behavior indicates that the DM interaction, although it may adversely affect the peak current, leads to an increase in the total charge over time. Furthermore, increasing \(\kappa\) from 0 to 2 significantly enhances \(Q\) at every point. This demonstrates that stronger anisotropy is a key factor in increasing the total charge.\\
These results reveal that the total charge is not solely determined by the peak current, but rather reflects the interplay between the peak height and the temporal width of the current pulse. The monotonic enhancement of \(Q\) with both \(J\) and \(D_z\) indicates that stronger spin interactions—despite their contrasting effects on the peak current—consistently improve the total charge extraction over time. This underscores the importance of considering time-integrated quantities when evaluating the performance of quantum photocells, as they capture the cumulative charge transfer that is directly relevant to practical device operation.

\subsection*{Appendix B: Coupling-to-Gap Ratio Analysis}

To assess the physical relevance of the coupling parameters used in the main text, we analyze the system in terms of the ratio of the couplings to a typical energy gap. As a reference, we use the qubit donor gap, \(\Delta E = E_{e_d} - E_{g_d}\), which is set to \(\Delta E = 1.9\) eV using the energy levels \(E_{g_d}=0.1\) eV, \(E_{e_d}=2.0\) eV, \(E_{g_D}=0.1\) eV, \(E_{e_D}=2.5\) eV, \(E_{u_D}=3.0\) eV, \(E_{g_a}=0.2\) eV, and \(E_{e_a}=1.75\) eV. This choice ensures that the gap is substantially larger than the coupling values considered in the ratio analysis, allowing the system to be characterized through the dimensionless ratios \(J/\Delta E\) and \(D_z/\Delta E\), rather than through the absolute magnitudes of \(J\) and \(D_z\).

\begin{figure}[!ht]
\centering
  \includegraphics[width=1\columnwidth]{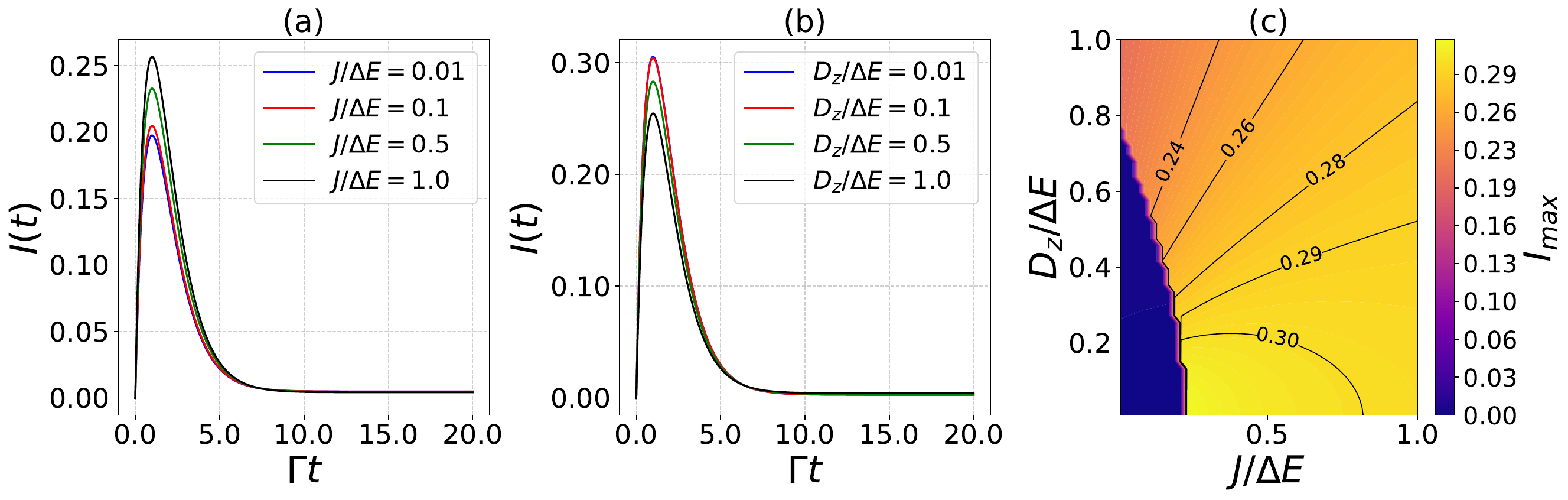}
  \caption{Time evolution of the transient photocurrent \(I(t)\) for different coupling-to-gap ratios, with the fixed coupling values set to \(\kappa=2\) and \(D_z=3\) for panel (a), and \(J=1\) and \(\kappa=2\) for panel (b). Panel (c) shows the contour plot of \(I_{max}\) as a function of \(J/\Delta E\) and \(D_z/\Delta E\). The ratios are varied from 0.01 to 1.0.}
\label{fig:ratio_large}
\end{figure}

Figure~\ref{fig:ratio_large} presents the time evolution of the transient photocurrent for different values of these ratios, with the remaining coupling parameters fixed at relatively large values (\(\kappa=2\) and \(D_z=3\) for panel a; \(J=1\) and \(\kappa=2\) for panel b). In panel (a), increasing \(J/\Delta E\) enhances the transient photocurrent, while in panel (b), increasing \(D_z/\Delta E\) suppresses it. These trends are fully consistent with the results reported in the main text. Panel (c) shows a contour plot of the maximum transient current \(I_{max}\) as a function of both \(J/\Delta E\) and \(D_z/\Delta E\). The contour plot reveals that \(I_{max}\) is maximized in the region where \(J/\Delta E\) is large and \(D_z/\Delta E\) is small, confirming the competing roles of the exchange and DM interactions.

\begin{figure}[!ht]
\centering
  \includegraphics[width=1\columnwidth]{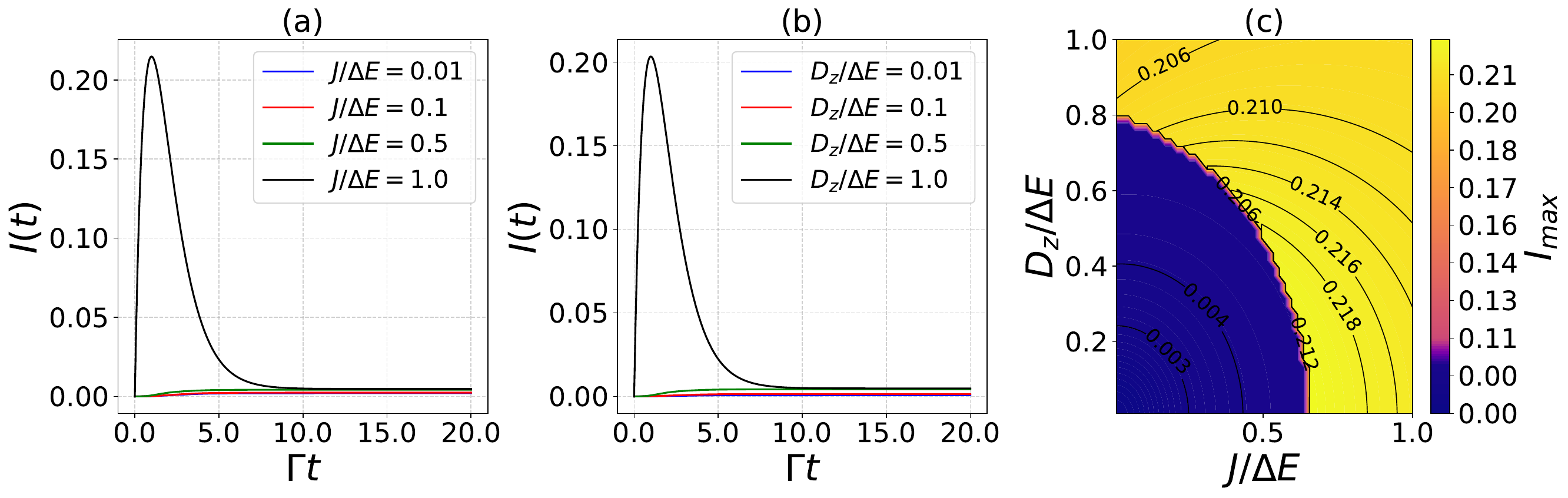}
  \caption{Time evolution of the transient photocurrent \(I(t)\) for different coupling-to-gap ratios, with the fixed coupling values set to \(\kappa=0.2\) and \(D_z=0.3\) for panel (a), and \(J=0.1\) and \(\kappa=0.2\) for panel (b). Panel (c) shows the contour plot of \(I_{max}\) as a function of \(J/\Delta E\) and \(D_z/\Delta E\). The ratios are varied from 0.01 to 1.0.}
\label{fig:ratio_small}
\end{figure}

Figure~\ref{fig:ratio_small} repeats the same analysis for a different set of fixed coupling values (\(\kappa=0.2\) and \(D_z=0.3\) for panel a; \(J=0.1\) and \(\kappa=0.2\) for panel b), which are smaller than those used in Fig.~\ref{fig:ratio_large}. In panel (a), increasing \(J/\Delta E\) still enhances the transient photocurrent, while in panel (b), increasing \(D_z/\Delta E\) also leads to an enhancement of the transient current, which is opposite to the behavior observed in Fig.~\ref{fig:ratio_large}(b). The contour plot in panel (c) further confirms that \(I_{max}\) is maximized in the region where both \(J/\Delta E\) and \(D_z/\Delta E\) are relatively large. In this regime, the transient photocurrent remains small unless the coupling-to-gap ratios exceed a certain threshold, indicating that a minimum coupling strength is required for a noticeable current.

Comparing Figs.~\ref{fig:ratio_large} and \ref{fig:ratio_small}, it becomes evident that the system behavior depends on the coupling regime. This regime-dependent behavior is a direct consequence of characterizing the system through the ratios \(J/\Delta E\) and \(D_z/\Delta E\), which capture the relative strength of the couplings with respect to the gap, rather than through the bare coupling values themselves.

\color{black}

\end{document}